\documentclass[letterpaper,10pt,twocolumn,superscriptaddress]{revtex4-1}
\usepackage{amsfonts,amssymb,amsmath,amsthm,graphicx}
\usepackage{url}
\usepackage{dsfont}
\usepackage{bbm}
\usepackage[usenames,dvipsnames]{xcolor}
 \usepackage{xcolor}
\usepackage{nicefrac}
\usepackage{natbib}
\usepackage{setspace}
\usepackage{subfigure}
\usepackage{tabularx,booktabs}
\usepackage{comment}
\usepackage[normalem]{ulem}
\usepackage{todonotes}
\usepackage{siunitx}
\usepackage{comment}
\usepackage{cancel}
\usepackage[colorlinks=true, citecolor=blue,urlcolor=blue,linkcolor=blue,filecolor=black]{hyperref}
\usepackage{filecontents}
\usepackage{csquotes}

\def\ket #1{\vert #1\rangle}

\newcommand{\beq}{\begin{equation}}
\newcommand{\eeq}{\end{equation}}

\begin{document}

\title{All-fiber autocompensating polarization encoder for Quantum Key Distribution}

 \author{Costantino Agnesi}
 \thanks{These authors contributed equally to this work.}
 \affiliation{Dipartimento di Ingegneria dell'Informazione, Universit\`a degli Studi di Padova, Via Gradenigo 6B - 35131 Padova, Italia}

\author{Marco Avesani}
\thanks{These authors contributed equally to this work.}
\affiliation{Dipartimento di Ingegneria dell'Informazione, Universit\`a degli Studi di Padova, Via Gradenigo 6B - 35131 Padova, Italia}

 \author{Andrea Stanco}
\affiliation{Dipartimento di Ingegneria dell'Informazione, Universit\`a degli Studi di Padova, Via Gradenigo 6B - 35131 Padova, Italia}
 
 \author{Paolo Villoresi}
\affiliation{Dipartimento di Ingegneria dell'Informazione, Universit\`a degli Studi di Padova, Via Gradenigo 6B - 35131 Padova, Italia}
 \affiliation{Istituto di Fotonica e Nanotecnologie - CNR, Via Trasea 7 - 35131 Padova, Italia}

 \author{Giuseppe Vallone}
 \email{vallone@dei.unipd.it}
\affiliation{Dipartimento di Ingegneria dell'Informazione, Universit\`a degli Studi di Padova, Via Gradenigo 6B - 35131 Padova, Italia}
 \affiliation{Istituto di Fotonica e Nanotecnologie - CNR, Via Trasea 7 - 35131 Padova, Italia}



\begin{abstract}
Quantum Key Distribution (QKD) allows distant parties to exchange cryptographic keys with unconditional security by encoding information on the degrees of freedom of photons.
Polarization encoding has been extensively used in QKD implementations along free-space, optical fiber and satellite-based links.
However, the polarization encoders used in such implementations are unstable, expensive, complex and can even exhibit side-channels that undermine the security of the implemented protocol.
Here we propose a self-compensating polarization encoder based on a Lithium Niobate phase modulator inside a Sagnac interferometer and implement it using only standard telecommunication commercial off-the-shelves components (COTS).
Our polarization encoder combines a simple design and high stability reaching an  intrinsic quantum bit error rate  as low as $0.2\%$.
Since realization is possible from the $800$ nm to the $1550$ nm band by using COTS, our polarization modulator is a promising solution for free-space, fiber and satellite-based QKD.
\end{abstract}

\maketitle

\section{Introduction}
Quantum key distribution (QKD) is an emerging quantum technology that allows two distant parties to distill a secret key with unconditional security by leveraging on the quantum mechanical nature of light \cite{Scarani2009}.
As several standard encryption schemes have been proven insecure and major steps have been made towards the development of the quantum computer  \cite{Wendin2017}, QKD has gained vast recognition. Indeed, the keys generated  in different terrestrial and space scenarios may  be used for  symmetric key cryptography, when high levels of privacy and long-term secrecy are required.
Several implementations of QKD systems have been reported in recent years, demonstrating the possibility of exploiting photonic degrees of freedom, such as polarization, time-bin and orbital angular momentum, in free-space, optical fiber or even satellite-based links \cite{Vallone2014,
vallone2016prl,Liao2017_daylight, Liao2017_satellite, Islam2017, Cozzolino2018, Boaron2018}. However, in-field adoption of QKD alongside the current telecommunication infrastructures requires all components of the system to exhibit a high degree of simplicity and stability. 

Widespread effort have been made to simplify the requirements of QKD systems and to enhance the stability of the practical implementations.
Recently, for example, a 3 state and 2 decoy state version of the BB84 protocol \cite{BB84} has been proposed \cite{Grunenfelder2018}, and demonstrated to be secure \cite{Rusca2018_APL, Rusca2018_PRA}, notably simplifying the requirements of the quantum state encoder and increasing the performances in the finite-key regime.
Likewise, a stable intensity modulator for decoy-state preparation \cite{Roberts2018}, as well as a stable phase modulator for time-bin encoding  \cite{Wang2018} have been demonstrated at repetition rates above GHz, both based on Sagnac interferometric configurations.

\begin{figure*}[t]
\centering
\fbox{\includegraphics[width=0.95\linewidth]{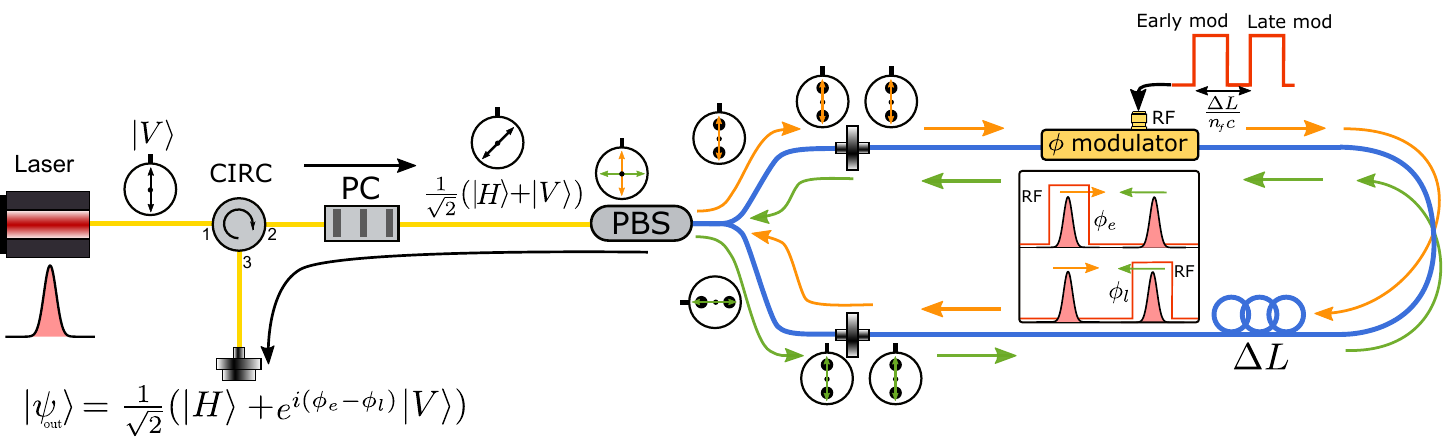}}
\caption{Schematic representation of the working principle of the POGNAC. SM fibers are represented in yellow and PM fibers in blue. For a detailed explanation see
section \ref{sec:to}.}
\label{fig:setup}
\end{figure*}

Despite polarization encoding being the predilected choice for free-space and satellite-based QKD experiments, few steps have been made to develop a simple and stable polarization state encoder.
The use of inline Lithium Niobate (LiNbO$_3$) modulators  has been an adopted solution \cite{Jofre2010,Grunenfelder2018}, where the birefringence of the crystal is controlled by an external RF field. The applied voltage changes the index of refraction of both polarization modes differently, introducing a relative phase between each polarization, thereby modulating the polarization state. However, high $\widetilde V_\pi$ voltage are needed to introduce a relative $\pi$ shift between orthogonal polarizations, usually a factor 1.5 higher when compared to $V_\pi$ of standard phase modulators.
Moreover, the stability of this inline configuration is critical, as the temperature variations caused by the environment or by the heating due to the RF internal power  induce drift in the resulting polarization state.

To address this problem, a double-pass autocompensating configuration with a Faraday Mirror has been proposed in \cite{Martinez2009}, which significantly improved long term stability. However, this approach has important drawbacks such as the use of non standard products (the polarization maintaining (PM) fiber has to be oriented at $45^\circ$ with respect to the optical axis of the LiNbO$_3$ crystal), high $\widetilde V_\pi$ voltages, the required use of high birefringence fibers to compensate for polarization mode dispersion and the need of Titanium-Diffused LiNbO$_3$ modulators able to guide two orthogonal polarizations that are hardly available at wavelength outside the C band. 
Moreover, any misalignment of the PM fiber with respect to the optical axis of the LiNbO$_3$ crystal will impact the possibility to generate orthogonal states.

Another approach is the use of four independent lasers which are then combined with polarization beamsplitters (PBSs), polarization controllers (PCs) and a beamsplitter (BS) \cite{Bacco2013,Vest2015, Liao2017_daylight, Liao2017_satellite}.
This approach, surely simplifies the electronic control of the QKD transmitter, but is expensive and power inefficient since it requires four times as many lasers, laser current drivers and temperature controllers.
Furthermore, the use of independent lasers could could open side-channels that undermine the security of the implementation in the presence of an eavesdropper.
In fact, differences in the temporal shapes and frequency spectrum of the independent laser pulses could be exploited to infer the polarization state without requiring a direct measurement \cite{Jofre2010}. 

In this letter we propose the POGNAC, a polarization modulator based on a LiNbO$_3$ phase modulator inside a Sagnac interferometer. We implement and test it by using standard off-the-shelf telecommunication components.
Our polarization modulator exhibits high degree of simplicity and stability, low intrinsic quantum bit error rate (QBER), and can be implemented for operation on both the $800$ nm band and the $1550$ nm band, rendering it compatible with free-space, optical fiber and satellite-based QKD.

\section{Theory of operation}
\label{sec:to}
Our proposed  polarization modulator based on a Sagnac interferometer (POGNAC) can be seen in Fig.\ref{fig:setup}.
A linearly polarized laser pulse enters the optical circulator (CIRC) in port 1 and exits in port 2.
A PC is then encountered which transforms the polarization state into $\ket{\psi} = \frac{1}{\sqrt{2}}(\ket{H} + e^{i\varphi_0} \ket{V} )$, a balanced superposition of horizontal and vertical polarization with arbitrary relative phase, i.e. any state on the equator of the Bloch sphere with $\ket H$ ($\ket V$) at the north (south) pole. 
The light is split into orthogonal linear polarizations by a fiber PBS.
It is important to note that each of the polarized beams exiting from the PBS is aligned to the slow axis of a PM fiber. This effectively maps the polarization degree of freedom onto the optical path of the photons, with the polarized light traveling along only the slow axis of the PM fibers of both PBS exit ports.
This is the standard behavior of fiber-based polarization beam combiners and splitters.

This PBS marks the beginning of the Sagnac interferometer, fully implemented with PM fibers.
The vertically polarized component travels in the clockwise direction (CW) while the horizontally polarized component travels in the counter-clockwise direction (CCW).
In the CW direction a LiNbO$_3$ phase modulator is first encountered introducing a phase $\phi_e$ to the CW propagating light pulse.
A PM fiber delay line is then encountered, after which the CW light pulse impinges once again on the PBS. 
The CW propagating light exits the Sagnac interferometer with horizontal polarization.
In the reverse direction, the CCW first encounters the PM fiber delay line. 
Then, the LiNbO$_3$ phase modulator which introduces a phase $\phi_\ell$ to the CCW propagating light pulse. 
Lastly, the CCW light pulse impinges once again on the PBS, exiting the Sagnac interferometer with vertical polarization.

Since inside the PM fiber Sagnac interferometer, both the CW and CCW travel along the fast axis of the PM fiber, no polarization mode dispersion is observed and a single polarization mode propagates in the phase modulator. This ensures that both CW and CCW pulses exit the Sagnac interferometer at the same time, perfectly recombining the two orthogonal polarization states after the PBS.
The emerging polarization state is thus given by 
\begin{equation}
 \ket{\psi^{\phi_e,\phi_\ell}_\mathrm{out}}  = \frac{1}{\sqrt{2}} \left[ \ket{H} + 
 e^{i(\phi_e-\phi_\ell- \varphi_0)} \ket{V} \right]\,.
\end{equation}

Since the CW pulse anticipates the arrival of the CCW pulse on the LiNbO$_3$ crystal by a factor  $\frac{\Delta L}{n_f c}$ (where $n_f$ is the index of refraction of the PM fibre and $c$ the velocity of light), by carefully timing the applied voltage on the phase modulator, the polarization state $\ket{\psi_\mathrm{out}}$ can be modulated. 
For sake of simplicity, lets suppose that $\varphi_0 = 0$.  
If no voltage (or equal voltage) is applied to the CW and CCW pulses, the polarization state remains unchanged, i.e.
\begin{equation}
\ket{\psi_\mathrm{out}^{0,0}} = \ket{D} = \frac{1}{\sqrt{2}}  \left[ \ket{H} + \ket{V} \right]. 
\end{equation}
Instead, if $V_{\pi/2}$ voltage is applied to the CW pulse and no voltage is applied to the CCW pulse, the output state becomes  
\begin{equation}
\ket{\psi_\mathrm{out}^{\frac{\pi}{2},0}} = \ket{L} = \frac{1}{\sqrt{2}}  \left[ \ket{H} + i \ket{V} \right].    
\end{equation}
Alternatively, if no voltage is applied to the CW pulse and $V_{\pi/2}$ voltage is applied to the CCW pulse
\begin{equation}
\ket{\psi_\mathrm{out}^{0,\frac{\pi}{2}}} = \ket{R} = \frac{1}{\sqrt{2}}  \left[ \ket{H} - i \ket{V} \right].   
\end{equation}
Finally, if  $V_{\pi}$  is applied to the CW (or CCW pulse), and no voltage is applied to the other, the output polarization state becomes
\begin{equation}
\ket{\psi_\mathrm{out}^{\pi,0}} = \ket{A} = \frac{1}{\sqrt{2}}  \left[ \ket{H} - \ket{V} \right].
\end{equation}
The modulated light pulses then exit through port 3 of the CIRC.

By noting that $\{ \ket{D}, \ket{A} \}$ and $\{ \ket{L}, \ket{R} \}$ form two mutually unbiased basis (MUBs), we can conclude that our proposed polarization modulator can generate the necessary polarization states to perform the standard BB84 QKD protocol \cite{BB84}.
We note that when $\varphi_0\neq 0$ the same scheme allows the generation of two MUBs lying on the equator of the Bloch sphere.
Furthermore, by choosing  $\{ \ket{L}, \ket{R} \}$ as the key generation states and $\ket{D}$ as the control state,  the simplified 3 polarization state version of BB84 \cite{Grunenfelder2018} can be implemented requiring only two voltage levels, i.e 0 and $V_{\pi/2}$, and fine positioning of the RF electrical pulse which can be done using digital outputs of a Field Programmable Gate Array (FPGA).

It can be useful to note that the four polarization states can also be generating by applying 4 different voltage levels, i. e. zero, $V_{\pi/2}$, $V_{\pi}$ and $V_{3\pi/2}$,   only to the CW or CCW pulse, always applying  zero voltage to the other.

\section{Experimental implementation}
\label{sec:exi}
We used a World Star Tech laser diode emitting light at $850$ nm  and an Hewlett-Packard 8013B pulse generator (PG) to generate laser pulses with  $1.2$ns FWHM duration. The light pulses first traversed a Glan-Thompson Polarizer, and was then coupled into a single mode (SM) fiber. In our implementation, the CIRC was replaced with a 50:50 BS. This replacement introduced additional 6dB of losses  which did not represent a problem since the light pulses were attenuated to the single photon level after the polarization modulator. A PC then transformed the polarization state into $\ket{\psi} = \frac{1}{\sqrt{2}} (\ket{H} + e^{i\varphi_0} \ket{V} )$. The light pulses then impinged a fiber based PBS. A $\Delta L = 1$m  PM fiber was used as the delay line inside the Sagnac interferometer. The RF signal used to drive the LiNbO$_3$ phase modulator were generated by an Avnet Zedboard FPGA board which was triggered by the PG. The FPGA generated squared pulses with $3$ns duration that could be arbitrarily delayed with respect to the trigger pulses with approximately $100$ps precision. This allowed us to send an electrical pulse that modulated either the CW propagating  or the CCW propagating pulse, or not to send any electrical pulse according to a previously established pseudorandom sequence. The electrical pulses were then  amplified to  $V_{\pi/2}$ by an RF amplifier and then sent to the phase modulator.
In this manner we simulated the polarization state transmission required by the simplified version of BB84 \cite{Grunenfelder2018}.
To test the generation of the $\ket{A}$ state, we replaced the FPGA with the Agilent 33521A arbitrary function generator that produced electrical pulses with $20$ns duration, allowing us to generate a $\ket{D}$, $\ket{A}$ sequence. This replacement was necessary to reach $V_\pi$ necessary to obtain the $\ket{A}$ state.
 
The light exited the polarization modulator through the 50:50 BS and then encountered an Optical Attenuator (OA) that attenuated to the single photon level. Then a PC is encountered transforming the states into $\ket{H}$, $\ket{V}$ and $\ket{D}$ states. The light pulses were then launched into free-space using a fiber collimator. A free-space polarization analyzer was then used to evaluate the performances of the polarization modulator. The analyzer was composed by a half-wave plate (HWP) and a PBS. This allowed us to measure in the $\{ \ket{H}, \ket{V} \}$ or in the $\{ \ket{D}, \ket{A} \}$ basis by simply rotating the HWP. The single photon detection was performed using Excelitas SPCM-AQRH single-photon avalanche diode and the quTAU timetagger. A computer was then used to analyze the results.

\begin{figure}[tbp]
\centering
\includegraphics[width=0.95\linewidth]{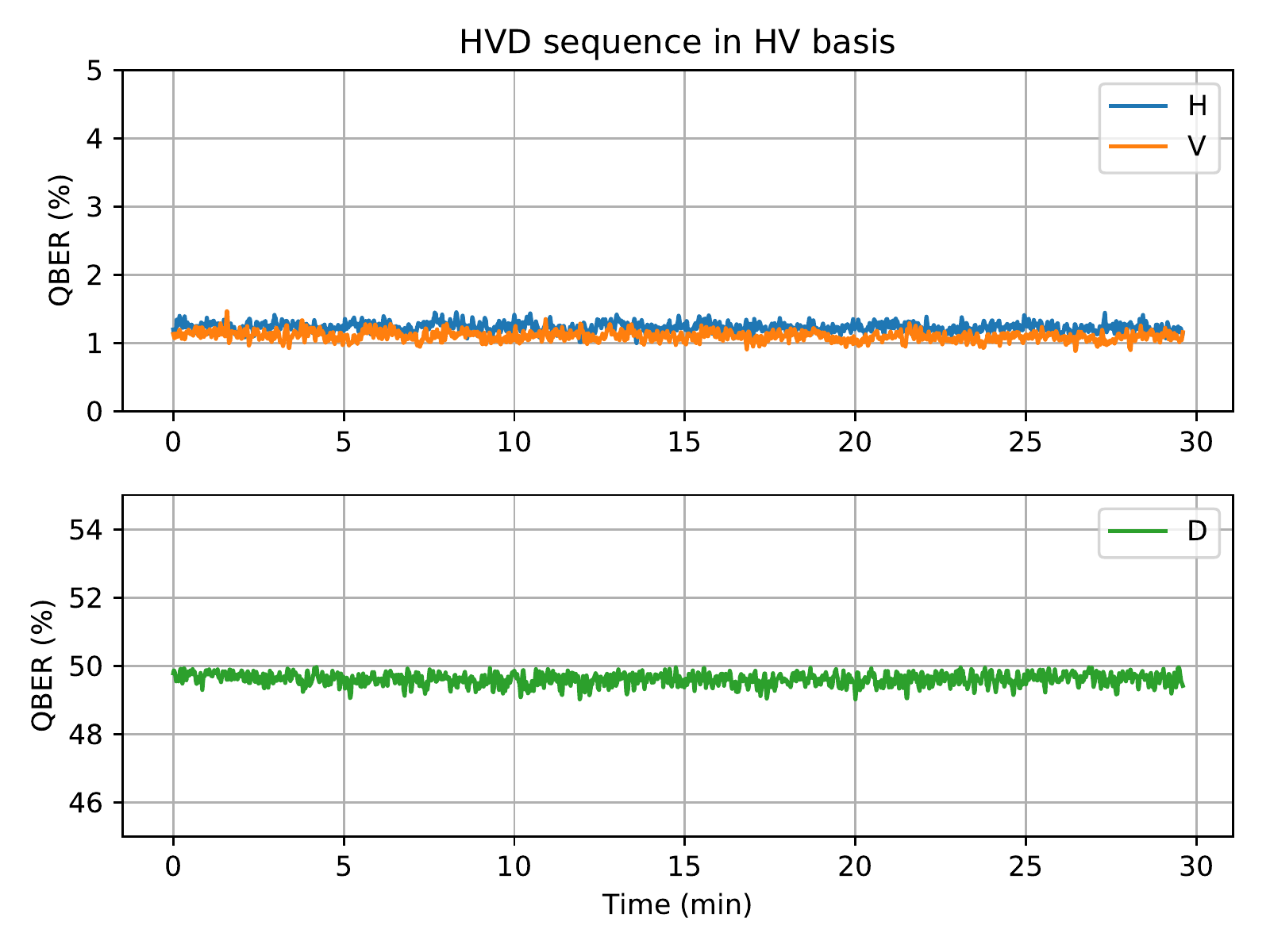}
\caption{QBER as a function of time for the pseudorandom $\{ \ket{H}, \ket{V}, \ket{D} \}$ sequence ($V_{\pi/2}$ modulation) measured in the $\{ \ket{H}, \ket{V} \}$ basis.}
\label{fig:HVD-HV}
\end{figure}
\section{Results}
\label{sec:res}
The pseudorandom $\{ \ket{H}, \ket{V}, \ket{D} \}$ sequence was continuously sent by our polarization encoder and measured by the free-space polarization analyzer in the $\{ \ket{H}, \ket{V} \}$ basis. Every three seconds, the data were analyzed by the computer and the QBER was calculated. The results can be seen in figure \ref{fig:HVD-HV}. An average QBER of $1.23 \pm 0.07\%$ was measured for $ \ket{H}$ and $1.10 \pm 0.07\%$ for $ \ket{V}$. Instead, for $\ket{D}$ a $49.6 \pm 0.2\%$ QBER was measured, as expected for a MUB state.

\begin{figure}[bhbp]
\centering
\includegraphics[width=0.95\linewidth]{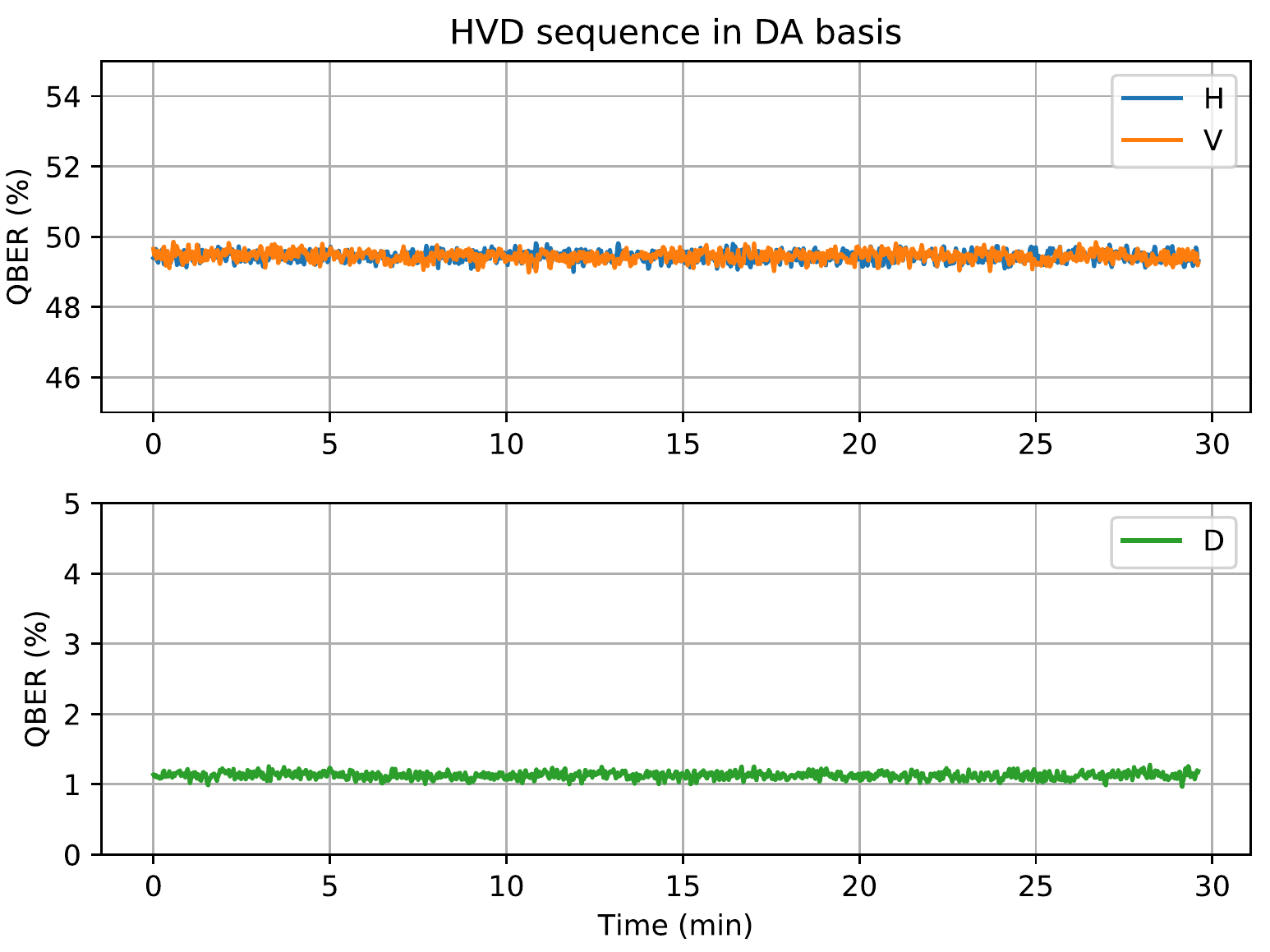}
\caption{QBER as a function of time for the pseudorandom $\{ \ket{H}, \ket{V}, \ket{D} \}$ sequence ($V_{\pi/2}$ modulation)  measured in the $\{ \ket{D}, \ket{A} \}$ basis.}
\label{fig:HVD-DA}
\end{figure}

After approximately 30 minutes, the HWP of the free-space polarization analyzer was rotated to measure in the $\{ \ket{D}, \ket{A} \}$ basis, without modifying the polarization encoder. As before, every  three seconds, the data were analyzed and the QBER was calculated. The results can be seen in figure \ref{fig:HVD-DA}.  An average QBER of $1.12 \pm 0.04\%$ was measured for $ \ket{D}$. Instead, for $\ket{H}$ and $\ket{V}$ a $49.4 \pm 0.1\%$ QBER was measured, as expected for MUB states.

\begin{figure}[tbp]
\centering
\includegraphics[width=0.95\linewidth]{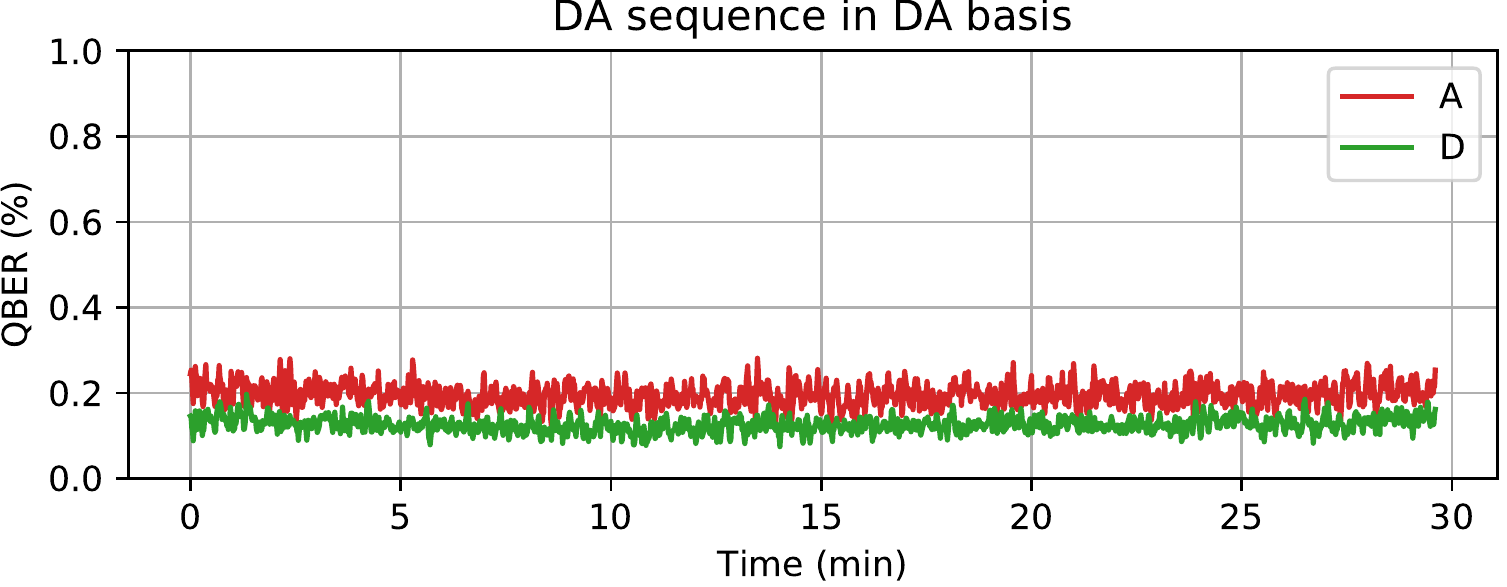}
\caption{QBER as a function of time for the $\{ \ket{D}, \ket{A} \}$ sequence ($V_\pi$ modulation) measured in the $\{ \ket{D}, \ket{A} \}$ basis.}
\label{fig:DA-DA}
\end{figure}

Similarly, to test the generation of the $\ket{A}$ state, a  $\{ \ket{D}, \ket{A} \}$ sequence was sent and the HWP of the free-space polarization analyzer was rotated to measure in the $\{ \ket{D}, \ket{A} \}$ basis.
Every three seconds, the data were analyzed and the QBER was calculated. The results can be seen in figure \ref{fig:DA-DA}.  An average QBER of $0.20 \pm 0.02\%$ was measured for $ \ket{A}$ and $0.13 \pm 0.1\%$ for $ \ket{D}$.
The QBER in this configuration is considerably lower if compared to the previous results. The reason can be attributed to the cleaner electrical RF pulses generated by the function generator respect to the ones generated by the FPGA.

\section{Conclusions}
In this paper we propose and experimentally test the POGNAC, a novel polarization encoder system for free-space, fiber and satellite QKD.
Compared to the previously proposed solutions our approach offers several key advantages.
The self-compensating Sagnac-loop design greatly improves long-term stability over inline implementations \cite{Jofre2010,Grunenfelder2018,Martinez2009}, making it insensible to temperature fluctuations and DC drifts.

Compared to the previously proposed autocompensating solution in \cite{Martinez2009} the POGNAC does not need a custom phase modulator. In fact in the Sagnac loop only one polarization is guided and standard Proton-Exchange Phase Modulators (PE-PM) can be used. The  Faraday Mirror solution instead requires Titanium-diffused phase modulators (TD-PM) that are able to support both polarizations. TD-PM are commercially available from few manufacturers, while PE-PM are standard telecom devices available at different wavelength, included the $800$ nm band, relevant for free space QKD.

Moreover, the modulation voltages required by our solution are considerably lower than previous proposals. In the POGNAC the phase modulation is directly converted in a polarization modulation. Instead in \cite{Jofre2010,Grunenfelder2018,Martinez2009} the applied voltage changes the index of refraction of both polarization modes differently, introducing a relative phase between each polarization, thereby modulating the polarization state. Usually, these implementation require a $V_{\pi}$ 1.5 times higher than our proposal.

Our experimental results show that low QBER can be obtained stably overtime without the need of an additional feedback system, greatly simplifying the design of a polarization QKD source. 
Such simplicity renders our source suitable for CubeSat missions, where a small footprint and low energy consumption are of critical importance \cite{Oi2017}. 
Furthermore, the temporal stability of the source attests the compatibility with QKD links with satellites even in Middle Earth Orbit \cite{Dequal2006}, or part of a Global Navigation Satellite Systems \cite{Calderaro2018}, where visibility times between the ground station and satellite can exceed the hour. 

Lastly, the configuration based on a Sagnac interferometer could allow for high repetition rates, up to few GHz, as recently demonstrated with an intensity modulator for decoy-state preparation \cite{Roberts2018}, as well as a stable phase modulator for time-bin encoding \cite{Wang2018}.

\bibliographystyle{apsrev4-1}
\bibliography{PolarizationSagnac}

\end{document}